\documentclass[seceq]{ptptex}
\usepackage{wrapft}
\usepackage{graphicx}

\newcommand{\bra}[1]{\langle {#1} |}     
\newcommand{\ket}[1]{| {#1} \rangle}     

\newcommand{\ovl}[1]{\overline{#1}}
\def\Box{\partial_\mu \partial^\mu}

\title{
Variational Approach to the Chiral Phase Transition\\
in the Linear Sigma Model 
}

\author{
Yasuhiko {\sc Tsue} and Kazuki {\sc Matsuda} 
}

\inst{
Physics Division, Faculty of Science, Kochi University, Kochi 780-8520, 
Japan
}

\recdate{
\today
}

\abst{
The chiral phase transition at finite temperature is investigated 
in the linear sigma model, which is regarded as a low energy effective 
theory of QCD with three momentum cutoff, 
in the variational method with the Gaussian approximation in 
the functional Schr\"odinger picture. 
It is shown that the Goldstone theorem is retained and the meson pair 
excitations are automatically included by taking into account the linear 
response to the external fields. 
It is pointed out that the behavior of chiral phase transition depends on 
the three-momentum cutoff, which leads to the careful treatment of 
the problem.
}

\begin{document}

\maketitle

\section{Introduction}

It is well known that the chiral symmetry plays an essential role 
in the quark and hadronic world governed by the quantum chromodynamics 
(QCD). 
Especially, the spontaneous chiral symmetry breaking leads to the 
existence of the pion,\cite{NAMBU} 
which plays an important role in the nuclear and hadron physics. 
One of the recent interests in the hadron physics is the chiral 
symmetry restoration in the context of the BNL 
Relativistic Heavy Ion Collider (RHIC) and/or the CERN Large Hadron Collider 
(LHC).\cite{QGP} 
It is believed that the chiral symmetry is restored in the high temperature 
and/or high baryon density, which might or may be created in the above 
mentioned experiments.

The finite temperature phase transition in the field theoretical treatment was 
investigated in the context of the phase transition in the early 
universe.\cite{Linde} 
Systematically, the field theoretical treatment for the systems at 
finite temperature was developed in the early stage.\cite{DJ,W}
Further, another approach\cite{EJP} 
to the finite temperature systems was devised with 
the variational technique in terms of 
the functional Schr\"odinger picture with Gaussian approximation\cite{JK,KV} 
for the bosonic field theory. 
In this paper, we intend to adopt a slight different variational approach 
to the bosonic field theory at finite temperature.

As for the low energy region of the strong interaction, the 
low energy effective theories of QCD give very useful tool until now 
to understand the physics of hadrons, especially, based on the 
chiral symmetry, while the lattice QCD simulation is developed 
for the problems related to the chiral symmetry. 
In the low energy region of QCD, the pion and its chiral partner, namely 
sigma meson, are important ingredients. 
As an effective theory of QCD, the linear sigma model\cite{GL} 
gives a powerful 
tool to understand the many properties of low energy hadron 
physics, which contains pions and sigma meson paying an attention 
to the chiral symmetry. 

In the linear sigma model, the chiral phase transition at finite temperature 
was firstly investigated in Ref.\citen{BG}, in which approximations 
developed in the many-body theory were used widely. 
However, according to the approximations, the order of the chiral phase 
transition was not settled definitely. 
Recently, in the context of the relativistic high energy heavy ion 
collision experiments, the renewed interests are pointed to the 
chiral phase transition in the O(N) linear sigma model from the 
theoretical sides.\cite{CH,RM} 
In the variational approach to the O(N) linear sigma model, 
it has been pointed out that the Goldstone theorem\cite{G} is not 
retained in naive treatments. 
For solving this problem, some works have been carried out.\cite{D,O,N,A} 
Also, one of the present authors (Y.T.) with Vautherin and Matsui gives 
another method to recover the Goldstone theorem in the variational 
approach to the O(N) linear sigma model, in which to take into account 
the linear response for the external fields is essential.\cite{TVM00}

As for the subject of the chiral phase transition at finite temperature, 
in Ref.\citen{CH}, so-called optimized perturbation theory is constructed 
for the O(N) linear sigma model, in which a certain resummation of the 
higher order terms is carried out. 
In the case $N=4$ for the realistic world, 
the Goldstone theorem is satisfied in this approach 
and it has been shown that 
this approach has revealed the second order chiral phase transition at 
finite temperature with realistic pion mass. 
On the other hand, the first order phase transition occurs with low 
pion mass including the case of the chiral limit. 
Another approach to the same subject was given in Ref.\citen{RM} 
which gave a self-consistent Hartree approximation for meson mass 
equations for one-loop effective potential. 
Both approaches are carried out the renormalization procedure 
because the O(N) linear sigma model is renormalizable.

As another approach, the large $N$ limit gives interesting features for 
O(N) linear sigma model. 
At zero temperature, the vacuum stability was first investigated.\cite{KK} 
For large $N$ limit in the O(N) linear sigma model, 
the Goldstone theorem is not broken. 
In the context of the chiral restoration, the large $N$ approach was 
investigated and the order of phase transition is discussed\cite{P} 
by using of the 
Cornwall-Jackiw-Tomboulis (CJT) effective potential.\cite{CJT}

In this paper, we investigate the chiral 
phase transition at finite temperature based on the time-dependent 
variational approach with Gaussian wave functional and with the 
imaginary time formalism of Green's functions,\cite{Matsubara} 
in which the linear responses 
for the external fields are taken into account. 
At zero temperature, it has already been shown that 
the Goldstone theorem is retained in our approach for the 
O(N) linear sigma model with $N=4$ mentioned above.\cite{TVM00} 
In this paper, we extend this approach to the finite temperature case 
and show the behavior of the chiral phase transition. 
We regard the O(4) linear sigma model as a low energy effective 
model of QCD, so we introduce the three momentum cutoff parameter $\Lambda$ 
which cuts the ultraviolet divergences of the meson loop appearing 
in our approach. 
It is shown that the pion-pion, pion-sigma meson and sigma-sigma mesons pair 
excitations are automatically included in our approach, 
which results to the Goldstone theorem and leads to 
the second order chiral phase transition with a reasonable parameterization in 
the O(4) linear sigma model.

This paper is organized as follows: 
In the next section, 
we recapitulate the basic ingredients of the variational approach 
with the linear response for the external fields to the O(N) linear sigma 
model. The pion and the sigma meson masses are given and it is shown 
that the meson pair excitations are automatically taken into account. 
In \S 3, we extend this approach to the finite temperature system 
in the Matsubara formalism.\cite{Matsubara} 
In \S 4, some numerical results and discussion are given. 
The last section is devoted to a summary and concluding remarks.

\section{Variational approach to the $O(4)$ linear 
sigma model satisfying the Goldstone theorem}

In this section, we recapitulate the basic ingredients of the variational 
method to the O(N) linear sigma model with the Gaussian wave functional 
and with the polarization tensors appearing in the linear response theory 
for the external fields.

\subsection{Variational approach and the breaking of the Goldstone theorem}

We summarize the time-dependent variational method following 
the Refs.\citen{TVM99} and \citen{TVM00}.

Let us start with the following Hamiltonian of the O(4) linear sigma 
model with the explicit symmetry breaking term:
\begin{eqnarray}\label{2-1}
& &{\cal H}({\mib x})=
{\cal H}_0({\mib x})-c\varphi_a({\mib x})\delta_{a0} \ , \nonumber\\
& &{\cal H}_0({\mib x})=\frac{1}{2}\pi_a({\mib x})\pi_a({\mib x})
+\frac{1}{2}\nabla\varphi_a({\mib x})\!\cdot\!\nabla\varphi_a({\mib x})
+\frac{1}{2}m_0^2\varphi_a({\mib x})\varphi_a({\mib x})
+\frac{\lambda}{24}(\varphi_a({\mib x})\varphi_a({\mib x}))^2 \ .
\nonumber
\\
& &
\end{eqnarray}
Here, $a$ runs from 0 to 3, where $\varphi_0$ means the sigma field operator 
and $\varphi_i$ with $i=1\sim 3$ mean the pion field operators. 
In this paper, we adopt the time-dependent variational approach in 
the functional Schr\"odinger picture. In this picture, the conjugate 
operators $\pi_a({\mib x})$ are replaced by the functional derivative 
$\pi_a({\mib x})=-i\delta/\delta \varphi_a({\mib x})$. 
A possible trial wave functional including the quantum fluctuations 
is a Gaussian wave functional which is written as 
\begin{equation}\label{2-2}
\Psi[\varphi]=\bra{\varphi}\Psi\rangle
={\cal N}e^{i\bra{\bar \pi}\varphi-{\bar \varphi}\rangle}
e^{-\bra{\varphi-{\bar \varphi}}\frac{1}{4G}
+i\Sigma\ket{\varphi-{\bar \varphi}}} \ , 
\end{equation}
where we used abbreviated notation as 
$
\bra{\bar \pi}\varphi\rangle=\int d^3{\mib x}\sum_{a=0}^{3}
{\bar \pi}_a({\mib x},t)\varphi_a({\mib x})
$
and\break
$ 
\bra{\varphi}G\ket{\varphi}=\int d^3{\mib x}d^3{\mib y}
\sum_{a,b=0}^3
\varphi_a({\mib x})G_{ab}({\mib x},{\mib y},t)\varphi_b({\mib y})
$. Also, ${\cal N}$ is a normalization factor. 
Here, the variational functions are ${\bar \varphi}_a({\mib x},t)$, 
${\bar \pi}_a({\mib x},t)$, $G_{ab}({\mib x}, {\mib y},t)$ and 
$\Sigma_{ab}({\mib x},{\mib y},t)$. 
In this approach, ${\bar \varphi}_a$ represent the mean fields and 
$G_{aa}$ represent the quantum fluctuations around the mean fields 
${\bar \varphi}_a$.  
The reason why we adopt the above form (\ref{2-2}) for the 
trial state with variational functions 
${\bar \varphi}_a({\mib x},t)$, 
${\bar \pi}_a({\mib x},t)$, $G_{ab}({\mib x}, {\mib y},t)$ and 
$\Sigma_{ab}({\mib x},{\mib y},t)$ is that the  
$({\bar \varphi}_a({\mib x},t), {\bar \pi}_a({\mib x},t))$ 
and 
$(G_{ab}({\mib x}, {\mib y},t), \Sigma_{ab}({\mib x},{\mib y},t))$ 
automatically satisfy the canonical variables condition\cite{MMSK} 
and/or canonicity conditions\cite{YK} in terms of the time-dependent 
Hartree-Fock theory developed in the nuclear many-body theory.

The time-dependence of the variational functions is governed by the 
following time-dependent variational principle: 
\begin{equation}\label{2-3}
\delta\int dt\bra{\Psi}i\frac{\partial}{\partial t}
-H\ket{\Psi} =0 \ ,
\end{equation}
where $H=\int d^3{\mib x}{\cal H}({\mib x})$. 
As a result, we derive a basic equation of motion for 
${\bar \varphi}_a({\mib x},t)$:\cite{TVM99} 
\begin{eqnarray}\label{2-4}
& &\left(\delta_{ab}\Box +m_{ab}^2(x)
-\frac{\lambda}{3}{\bar \varphi}_a{\bar \varphi}_b
\right)
{\bar \varphi}_b=c\delta_{a0} \ , 
\end{eqnarray}
where we can eliminate ${\bar \pi}$ due to the honor of the 
canonicity condition for ${\bar \varphi}$ and ${\bar \pi}$. 
Here, the ``mass" function $m_{ab}^2(x)$ are defined by 
\begin{eqnarray}\label{2-4-2}
& &m_{ab}^2(x)=\left(
m_0^2+\frac{\lambda}{6}{\bar \varphi}^2+\frac{\lambda}{6}
{\rm tr}\ S^{(0)}(x,x)\right)\delta_{ab}
+\frac{\lambda}{3}{\bar \varphi}_a(x){\bar \varphi}_b(x)
+\frac{\lambda}{3}S_{ab}^{(0)}(x,x)\ . \qquad
\end{eqnarray}
The function $S^{(0)}_{ab}(x,y)$ corresponds 
to the propagator and is formally obtained as 
\begin{eqnarray}\label{2-5}
& &S^{(0)}(x,y)=\int\frac{d^4p}{(2\pi)^4}S^{(0)}(p)e^{ip(x-y)} \ , 
\nonumber\\
& &S^{(0)}(p)=\frac{i}{-p^2+m^2(x)-i\epsilon}
\end{eqnarray}
The solution of the variational equation for $G_{aa}({\mib x},{\mib x},t)$ 
is given in terms of the diagonal element of the above propagator as 
\begin{eqnarray}\label{2-7}
G_{aa}({\mib x}, {\mib x}, t)\equiv 
G(m_{aa}^2)&=&S_{aa}^{(0)}(x,x)\ .
\end{eqnarray}

In the Hamiltonian, the direction of the spontaneous chiral 
symmetry breaking is taken in the sigma direction. Thus, the solution of 
the mean filed ${\bar \varphi}$ is assumed\break
as
\footnote{
Here, the first component of ${\bar \varphi}^{(0)}$, $\varphi_0$, 
in (\ref{2-8}) is not a 
field operator appearing in (\ref{2-1}). 
We think that there is no confusion. 
} 
\begin{equation}\label{2-8}
{\bar \varphi}={\bar \varphi}^{(0)}=
\left(
\begin{array}{@{\,}c@{\,}}
\varphi_0 \\
0 \\
0\\
0
\end{array}
\right) \ . 
\end{equation}
Then, 
we define the sigma meson mass $M$ and the pion mass $\mu$, 
which correspond to the masses derived in the Hartree approximation, as 
\begin{eqnarray}\label{2-9}
& &M^2=m_{00}^2=m_0^2+\frac{\lambda}{2}\varphi_0^2+\frac{\lambda}{2}G(M^2)
+\frac{\lambda}{2}G(\mu^2) \ , \nonumber\\
& &\mu^2=m_{11}^2=m_{22}^2=m_{33}^2=
m_0^2+\frac{\lambda}{6}\varphi_0^2+\frac{\lambda}{6}G(M^2)
+\frac{5}{6}\lambda G(\mu^2) \ .
\end{eqnarray}
From (\ref{2-4}) in the static case, we can determine the chiral condensate 
$\varphi_0$ by means of the mean filed approximation:
\begin{equation}\label{2-10}
\left(M^2-\frac{\lambda}{3}\varphi_0^2\right)\varphi_0=c \ . 
\end{equation}
The solution of the above equation is given in the two phases up to the 
order of $c$: 
\begin{eqnarray}
& &
\frac{\lambda}{3}\varphi_0^2 \approx M^2-\sqrt{\frac{\lambda}{3}}\frac{c}{M}
\ , \qquad (\hbox{\rm chiral\ broken\ phase}) 
\label{2-11}\\
& &\varphi_0 \approx \frac{c}{M^2}
\ . \qquad (\hbox{\rm chiral\ symmetric\ phase}) 
\label{2-12}
\end{eqnarray}
It should be noted here that the mass $\mu$ which corresponds to the pion 
mass does not satisfy the Goldstone theorem in the chiral limit, 
namely, even if $c=0$, then $\mu \neq 0$.

\subsection{Pion mass and recovery of the Goldstone theorem}

To see that the Goldstone theorem is satisfied in the variational approach, 
we introduce the external source field $J_a$ coupled with the $\varphi_a$: 
\begin{equation}\label{2-13}
{\cal H}'= {\cal H}-\sum_{a}J_a\varphi_a \ .
\end{equation}
By this source term $J_a\varphi_a$, the various quantities obtained in 
the previous subsection are modified as 
follows: 
\begin{eqnarray}\label{2-14}
& & {\bar \varphi}\longrightarrow {\bar \varphi}+\delta\varphi \ , \nonumber\\
& & m^2(x)\longrightarrow m^2(x)+\delta m^2 \ , \nonumber\\
& & S(x,x)\longrightarrow S^{(0)}(x,x)+\delta S(x,x) \ .
\end{eqnarray}
As a result, the equation of motion for ${\bar \varphi}$ is recast into the 
equation of motion for $\delta \varphi$ as 
\begin{eqnarray}\label{2-15}
& &\left[
-q^2\delta_{ab}+m_{ab}^2(x)-\frac{\lambda}{3}{\bar \varphi}_a^{(0)}\times
{\bar \varphi}_b^{(0)}\right]\delta\varphi_b
+\delta m_{ab}^2(x){\bar \varphi}_b^{(0)}
\nonumber\\
& &\qquad\qquad\qquad\qquad
-\frac{\lambda}{3}\left(\delta\varphi_a\times{\bar \varphi}_b^{(0)}
+{\bar \varphi}_a^{(0)}\times \delta\varphi_b\right){\bar \varphi}_b^{(0)}
=J_a \ , 
\end{eqnarray}
where ${\bar \varphi}_a^{(0)}$ is a solution of Eq.(\ref{2-4}) and 
\begin{equation}\label{2-16}
\delta m_{ab}^2(x)
=\left(\frac{\lambda}{3}\varphi_0\delta\varphi_0+\frac{\lambda}{6}
{\rm tr}\ \delta S\right)\delta_{ab}
+\frac{\lambda}{3}\left({\bar \varphi}_a\times \delta\varphi_b +
\delta\varphi_a\times{\bar \varphi}_b\right)
+\frac{\lambda}{3}\delta S_{ab} \ .
\end{equation}

Hereafter, we consider the case that 
the source field $J_a$ has the following form: 
\begin{equation}\label{2-17}
J_a(x)=\varepsilon\delta_{ai}e^{i(\omega t-{\bf q}\cdot{\bf x})} \ .
\end{equation}
Then, the shifted quantities $\delta m^2$ and $\delta S$ as well as 
$\delta \varphi$ have the same time- and coordinate-dependence as $J_a$:
\begin{eqnarray}
& &\delta m_{ab}^2(x)=\delta{\hat m}_{ab}^2 e^{i(\omega t-{\bf q}\cdot{\bf x})}
\ , 
\label{2-18}\\
& &\delta S_{ab}(x,x)=\delta{\hat m}_{ab}^2  
\Pi_{ab}^{({\rm pol})}(q^2)e^{i(\omega t-{\bf q}\cdot{\bf x})} 
\ . 
\label{2-19}
\end{eqnarray}
Here, we have defined the polarization tensor $\Pi_{ab}^{({\rm pol})}(q^2)$ 
as 
\begin{eqnarray}\label{2-20}
& &
\Pi_{ab}^{({\rm pol})}(q^2)=
-\int\frac{d^4 p}{i(2\pi)^4}S_{aa}^{(0)}(p+q)S_{bb}^{(0)}(p) \ .
\end{eqnarray}

Here, let us consider the pion mass which should satisfy the Goldstone theorem 
in the chiral limit. In order to derive the pion mass in the variational 
approach with external fields, 
we investigate the response of the following external source field: 
\begin{equation}\label{2-21}
J_a(x)=\varepsilon\delta_{a1}e^{i(\omega t-{\bf q}\cdot{\bf x})} \ . 
\end{equation}
For this source field, only $\delta \varphi_1$ responds, namely, 
\begin{equation}\label{2-22}
\delta\varphi_1 \neq 0 \ , \qquad \delta\varphi_a=0\quad (a=0,2,3) \ .
\end{equation}
We take an ansatz for 
the structure of matrices $\delta m_{ab}$ and $\delta S_{ab}$: 
\begin{equation}\label{2-23}
\delta m_{ab}^2(x)=
\left(
\begin{array}{@{\,}cccc@{\,}}
0 & \eta & 0 & 0 \\
\eta & 0 & 0 & 0 \\
0 & 0 & 0 & 0 \\
0 & 0 & 0 & 0
\end{array}
\right)
\ , \qquad
\delta S_{ab}(x)=
\left(
\begin{array}{@{\,}cccc@{\,}}
0 & \delta S & 0 & 0 \\
\delta S & 0 & 0 & 0 \\
0 & 0 & 0 & 0 \\
0 & 0 & 0 & 0
\end{array}
\right)
\ .
\end{equation}
It will be understood later that this ansatz is correct.

\begin{figure}[t]
\begin{center}
\includegraphics[height=3cm]{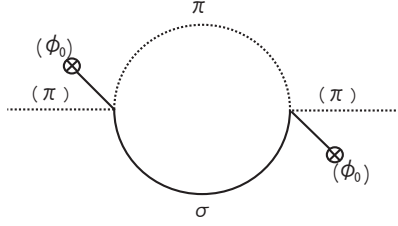}
\caption{The polarization tensor $\Pi_{01}^{({\rm pol})}$ is depicted 
diagrammatically. Here, $\Pi_{01}^{({\rm pol})}$ means a part which is 
gotten rid of the external line $(\pi)$ and the chiral condensate 
${\varphi}_0$ $(\phi_0)$ in the above diagram. 
}
\label{fig:2-1}
\end{center}
\end{figure}

For $a=1$, we derive the equation of motion from (\ref{2-15}) as 
\begin{eqnarray}\label{2-24}
& &(-q^2+\mu^2)\delta\varphi_1 + \delta m_{10}^2\varphi_0
-\frac{\lambda}{3}\varphi_0^2\delta\varphi_1 = J_1\ , 
\end{eqnarray}
where 
\begin{eqnarray}\label{2-25}
& &\delta m_{10}^2=\frac{\lambda}{3}\varphi_0\delta\varphi_1
+\frac{\lambda}{3}\delta S_{10} \ .
\end{eqnarray}
From (\ref{2-19}) and (\ref{2-25}), we obtain 
\begin{equation}\label{2-26}
\delta m_{10}^2\varphi_0=
\frac{\lambda}{3}\varphi_0^2\cdot 
\frac{1}{1-\frac{\lambda}{3}\Pi_{01}^{({\rm pol})}(q^2)}
\cdot\delta\varphi_1 \ .
\end{equation}
Thus, from (\ref{2-24}) and the above relation, we can derive the 
equation of motion for $\delta \varphi_1$ under the source field $J_1$:
\begin{equation}\label{2-27}
\left(-q^2+\mu^2+
\frac{\lambda}{3}\varphi_0^2\cdot 
\frac{\frac{\lambda}{3}\Pi_{01}^{({\rm pol})}(q^2)}
{1-\frac{\lambda}{3}\Pi_{01}^{({\rm pol})}(q^2)}
\right)\delta\varphi_1=J_1 \ .
\end{equation}
The full propagator $S_{ab}(x,y)$ is thus defined as 
\begin{equation}\label{2-28}
\delta\varphi_a(x)/\delta J_b(y)=S_{ab}(x,y) \ .
\end{equation}
Then, the mass is defined as the pole of the full propagator: 
\begin{equation}\label{2-29}
M_{\pi}^2=\mu^2+\frac{\lambda}{3}\varphi_0^2\cdot 
\frac{\frac{\lambda}{3}\Pi_{01}^{({\rm pol})}(0)}
{1-\frac{\lambda}{3}\Pi_{01}^{({\rm pol})}(0)}
\end{equation}
This mass corresponds to the pion mass. 
From Eq.(\ref{2-20}), the polarization tensor $\Pi_{01}^{({\rm pol})}(q^2)$ 
appearing in Eq.(\ref{2-27}) or (\ref{2-29}) reveals the pion and the sigma 
meson pair excitation, which is shown in Fig.{\ref{fig:2-1}} diagrammatically. 

Let us try to derive more explicit form for the pion mass $M_{\pi}$. 
From (\ref{2-7}) and (\ref{2-5}), the following relation is satisfied:
\begin{equation}\label{2-30}
G(M^2)-G(\mu^2)=\int \frac{d^4 p}{(2\pi)^4}
\left(\frac{i}{-p^2+M^2-i\epsilon}-\frac{i}{-p^2+\mu^2-i\epsilon}\right)
\end{equation}
Thus, the polarization tensor $\Pi_{01}^{({\rm pol})}$ can be expressed as 
\begin{equation}\label{2-31}
\Pi_{01}^{({\rm pol})}(0)=\frac{G(M^2)-G(\mu^2)}{M^2-\mu^2} \ . 
\end{equation}
From Eqs.(\ref{2-9}) and (\ref{2-11}) or (\ref{2-12}), 
the following relations 
are derived easily: 
\begin{eqnarray}
& &\mu^2=\sqrt{\frac{\lambda}{3}}\frac{c}{M}-\frac{\lambda}{3}
(G(M^2)-G(\mu^2))
\ , \qquad (\hbox{\rm chiral\ broken\ phase}) 
\label{2-32}\\
& &\mu^2=M^2-\frac{\lambda c^2}{3M^4}-\frac{\lambda}{3}
(G(M^2)-G(\mu^2))
\ . \qquad (\hbox{\rm chiral\ symmetric\ phase})
\label{2-33}
\end{eqnarray}
Thus, we obtain 
\begin{eqnarray}
& &\left(1-\frac{\lambda}{3}\Pi_{01}^{({\rm pol})}(0)\right)\mu^2=
-\frac{\lambda}{3}M^2\Pi_{01}^{({\rm pol})}(0)
+\sqrt{\frac{\lambda}{3}}\frac{c}{M}
\ , \nonumber\\
& &\qquad\qquad\qquad\qquad\qquad\qquad\qquad\qquad\qquad\qquad 
(\hbox{\rm chiral\ broken\ phase}) 
\label{2-34}\\
& &\left(1-\frac{\lambda}{3}\Pi_{01}^{({\rm pol})}(0)\right)\mu^2=
\left(1-\frac{\lambda}{3}\Pi_{01}^{({\rm pol})}(0)\right)M^2
-{\frac{\lambda}{3}}\frac{c^2}{M^4} \ . \nonumber\\
& &\qquad\qquad\qquad\qquad\qquad\qquad\qquad\qquad\qquad\qquad 
(\hbox{\rm chiral\ symmetric\ phase}) 
\label{2-35}
\end{eqnarray}
Substituting the above relation into (\ref{2-29}) and using 
(2$\cdot$11) or (2$\cdot$12) again, we finally obtain the pion mass in the 
chiral symmetry broken and chiral symmetric phases as 
\begin{eqnarray}
& &M_{\pi}^2=\sqrt{\frac{\lambda}{3}}\frac{c}{M}
\ , \qquad (\hbox{\rm chiral\ broken\ phase}) \\
& &M_{\pi}^2=M^2-\frac{\lambda c^2}{3M^4}
\ , \qquad (\hbox{\rm chiral\ symmetric\ phase}) \ . 
\end{eqnarray}
As is expected, in the chiral broken phase, if we take the chiral limit 
$c=0$, the pion mass is reduced to zero which realizes the Goldstone 
theorem.\cite{TVM00} 
As is seen in the polarization tensor $\Pi_{01}^{({\rm pol})}$ in 
Eq.(\ref{2-20}) or in Fig.\ref{fig:2-1}, 
the real pion mass contains the effects of the pion and the sigma meson 
pair excitations. 
Also, as is seen from the second term 
of the right-hand side in Eq.(\ref{2-29}), the polarization tensor 
appears in the denominator, so this meson pair excitation is 
continuously connected as 
$1/(1-\lambda \Pi_{01}^{({\rm pol})}/3)=1+\lambda \Pi_{01}^{({\rm pol})}/3
+\lambda \Pi_{01}^{({\rm pol})}/3\cdot \lambda \Pi_{01}^{({\rm pol})}/3
+\cdots$. 
Thus, the meson pair excitation is automatically included in our approach 
and leads to the consistency for the Goldstone theorem.

\subsection{Sigma meson mass}

\begin{figure}[t]
\begin{center}
\includegraphics[height=3.5cm]{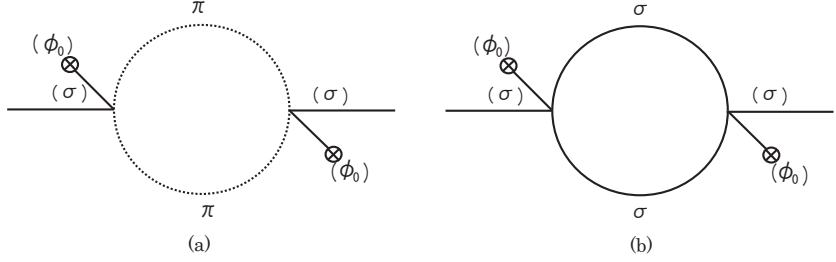}
\caption{The polarization tensor $\Pi_{11}^{({\rm pol})}$ 
and $\Pi_{00}^{({\rm pol})}$ are depicted 
diagrammatically in (a) and (b), respectively. 
Here, $\Pi_{11}^{({\rm pol})}$ and $\Pi_{00}^{({\rm pol})}$ 
mean parts which are 
gotten rid of the external line $(\sigma)$ and the chiral condensate 
${\varphi}_0$ $(\phi_0)$ in the above diagrams. 
}
\label{fig:2-2}
\end{center}
\end{figure}

In this subsection, we derive the sigma meson mass as is similar to 
deriving the pion mass in the previous subsection. 
In order to derive the sigma meson mass, we need to introduce the external 
source field for the sigma direction
\begin{equation}\label{2-38}
J_a(x)=\varepsilon\delta_{a0}e^{i(\omega t-{\bf q}\cdot{\bf x})} \ , 
\end{equation}
and to see the response of this external field. 
In this case, only $\delta {\varphi}_0$ responds: 
\begin{equation}\label{2-39}
\delta\varphi_0 \neq 0 \ , \qquad \delta\varphi_a=0\quad (a=1,2,3) \ .
\end{equation}
We assume the structure of matrices as 
\begin{equation}\label{2-40}
\delta m_{ab}^2(x)=
\left(
\begin{array}{@{\,}cccc@{\,}}
\zeta & 0 & 0 & 0 \\
0 & \eta' & 0 & 0 \\
0 & 0 & \eta' & 0 \\
0 & 0 & 0 & \eta'
\end{array}
\right)
\ , \qquad
\delta S_{ab}(x)=
\left(
\begin{array}{@{\,}cccc@{\,}}
\delta S_{\sigma} & 0 & 0 & 0 \\
0 & \delta S_{\pi} & 0 & 0 \\
0 & 0 & \delta S_{\pi} & 0 \\
0 & 0 & 0 & \delta S_{\pi}
\end{array}
\right)
\ .
\end{equation}
For $a=0$, we obtain the equation of motion for $\delta \varphi_0$ as 
\begin{eqnarray}\label{2-41}
& &(-q^2+M^2-\lambda\varphi_0^2)\delta\varphi_0 + \delta m_{00}^2\varphi_0
= J_0\ , 
\end{eqnarray}
where 
\begin{eqnarray}
& &\delta m_{00}^2={\lambda}\varphi_0\delta\varphi_0
+\frac{\lambda}{2}\delta S_{\sigma}+\frac{\lambda}{2}\delta S_{\pi} \ , 
\label{2-42}\\
& &\delta m_{11}^2=\delta m_{22}^2=\delta m_{33}^2=
\frac{\lambda}{3}\varphi_0\delta\varphi_0
+\frac{\lambda}{6}\delta S_{\sigma}+\frac{5}{6}\lambda\delta S_{\pi} \ . 
\label{2-43}
\end{eqnarray}
From (\ref{2-41}) and (\ref{2-19}), we obtain
\begin{equation}\label{2-44}
\delta m_{00}^2\varphi_0=
\frac{1+
\frac{\lambda}{6}{\ovl \Pi_{11}(q^2)}}{
1-\frac{\lambda}{2}\Pi_{00}^{({\rm pol})}(q^2)\left(
1+\frac{\lambda}{6}{\ovl \Pi}_{11}(q^2)\right)}
\cdot\lambda\varphi_0^2\delta\varphi_0 \ ,
\end{equation}
where we define the reduced polarization tensor as 
\begin{equation}\label{2-45}
{\ovl \Pi}_{11}(q^2)=\frac{\Pi_{11}^{({\rm pol})}(q^2)}{1-\frac{5}{6}
\lambda\Pi_{11}^{({\rm pol})}(q^2)} \ .
\end{equation}
As a result, we can obtain the following relation from (\ref{2-41}): 
\begin{equation}\label{2-46}
\left(-q^2+M^2+
{\lambda}\varphi_0^2\cdot 
\frac{\frac{\lambda}{2}\Pi_{00}^{({\rm pol})}(q^2)\left(
1+\frac{\lambda}{6}{\ovl \Pi}_{11}(q^2)\right)+
\frac{\lambda}{6}{\ovl \Pi}_{11}(q^2)}
{1-\frac{\lambda}{2}\Pi_{00}^{({\rm pol})}(q^2)\left(
1+\frac{\lambda}{6}{\ovl \Pi}_{11}(q^2)\right)}
\right)\delta\varphi_0=J_0\ .
\end{equation}
Thus, we can derive the sigma meson mass $M_\sigma$ as a pole of the 
full propagator $\delta \varphi_0/\delta J_0$: 
\begin{equation}\label{2-47}
M_{\sigma}^2=M^2
+{\lambda}\varphi_0^2\cdot 
\frac{\frac{\lambda}{2}\Pi_{00}^{({\rm pol})}(0)\left(
1+\frac{\lambda}{6}{\ovl \Pi}_{11}(0)\right)
+\frac{\lambda}{6}{\ovl \Pi}_{11}(0)}
{1-\frac{\lambda}{2}\Pi_{00}^{({\rm pol})}(0)\left(1+\frac{\lambda}{6}
{\ovl \Pi}_{11}(0)\right)} \ .
\end{equation}
In this formula, the polarization tensors $\Pi_{11}^{({\rm pol})}$ and 
$\Pi_{00}^{({\rm pol})}$ are automatically included. 
Here, 
$\Pi_{11}^{({\rm pol})}$ and $\Pi_{00}^{({\rm pol})}$ represent 
the pion-pair and the sigma meson-pair excitation, respectively, 
as is depicted in Fig.\ref{fig:2-2}.

\section{Pion and sigma meson masses at finite temperature}

In this section, we investigate the pion and sigma meson mass at finite 
temperature in this variational approach. 
To extend the method developed in the previous section 
to the finite temperature case, 
the $p_0$ integral is simply replaced to the Matsubara sum: 
\begin{equation}\label{3-1}
\int\frac{d^4 p}{(2\pi)^4} f(p)\longrightarrow 
iT\sum_{n=-\infty}^{\infty}
\int \frac{d^3{\mib p}}{(2\pi)^3}f(i\omega_n,{\mib p}) \ . 
\end{equation}
Here, $\omega_n$ is the Matsubara frequency of the bosons:
\begin{equation}\label{3-2}
\omega_n=2n\pi T \ . \qquad (n:{\rm integer})
\end{equation}
Under this extension, the diagonal element of the propagator is written as 
\begin{eqnarray}\label{3-3}
& &S(x,x)=iT\sum_{n=-\infty}^{\infty}
\int \frac{d^3{\mib p}}{(2\pi)^3}S(i\omega_n,{\mib p}) \ , 
\end{eqnarray}
where the Fourier transform $S(i\omega_n, {\mib p})$ is given as 
\begin{eqnarray}\label{3-4}
& &S(i\omega_n,{\mib p})=\frac{i}{\omega_n^2+{\mib p}^2+m^2(x)-i\epsilon} \ .
\end{eqnarray}
Also, the polarization tensor (\ref{2-20}) is rewritten as 
\begin{eqnarray}\label{3-5}
& &\Pi_{ab}^{({\rm pol})}(i\omega_n,{\mib q})
=-T\sum_{m=-\infty}^{\infty}
\int\frac{d^3{\mib p}}{(2\pi)^3}
S_{aa}(i\omega_m+i\omega_n,{\mib p}+{\mib q})S_{bb}(i\omega_m,{\mib p})\ . 
\end{eqnarray}

Using the well-known technique of the contour integral, the sum of the 
Matsubara frequency is easily taken and the resultant expression for $S(x,x)$ 
can simply expressed in terms of the bose distribution function $n^{(\pm)}$ 
as 
\begin{eqnarray}
& &S(x,x)=\frac{1}{4\pi^2}\int_0^{\infty}
d|{\mib p}|\frac{{\mib p}^2}{\sqrt{{\mib p}^2+m^2(x)}}
\left(n^{(+)}-n^{(-)}\right)\ , 
\label{3-6}\\
& &n^{(\pm)}
=\frac{1}{\exp\left(\pm\frac{\sqrt{{\mib p}^2+m^2(x)}}{T}\right)-1} \ . 
\label{3-7}
\end{eqnarray}
Similarly, the polarization tensors at finite temperature are calculated and 
the results are given as follows: 
\begin{eqnarray}
& &\Pi_{aa}^{({\rm pol})}(0,{\mib 0})
=-\frac{1}{8\pi^2}\int_0^{\Lambda}d|{\mib p}|
\frac{{\mib p}^2}{{\mib p}^2+m_a^2(x)} \nonumber\\
& &\qquad\qquad\qquad
\times
\biggl[\frac{1}{T}\left((n_a^{(+)})^2e^{\sqrt{{\mib p}^2+m_a(x)^2}/T}
+(n^{(-)}_a)^2e^{-\sqrt{{\mib p}^2+m_a(x)^2}/T}\right)\nonumber\\
& &\qquad\qquad\qquad\qquad
+
\frac{1}{\sqrt{{\mib p}^2+m_a^2(x)}}(n_a^{(+)} - (n^{(-)}_a)\biggl] \ , 
\label{3-8}
\\
& &\Pi_{ab}^{({\rm pol})}(0,{\mib 0})
=\frac{1}{4\pi^2(m_a^2-m_b^2)}\int_0^{\Lambda}d|{\mib p}|
{\mib p}^2\nonumber\\
& &\qquad\qquad\qquad
\times\biggl[\frac{1}{\sqrt{{\mib p}^2+m_a^2(x)}}
(n_a^{(+)}-n^{(-)}_a)-
\frac{1}{\sqrt{{\mib p}^2+m_b^2(x)}}(n_b^{(+)} - (n^{(-)}_b)\biggl] \ , 
\nonumber\\
& &
\qquad\qquad\qquad\qquad\qquad\qquad\qquad\qquad
\qquad\qquad\qquad\qquad\qquad\qquad
(a\neq b)
\label{3-9}
\end{eqnarray}
where we define $m_a=m_{aa}$ and $n_a=n_{aa}$ for simplicity. 

%

\section{Numerical results and discussion}

\begin{figure}[t]
\begin{center}
\includegraphics[height=5.3cm]{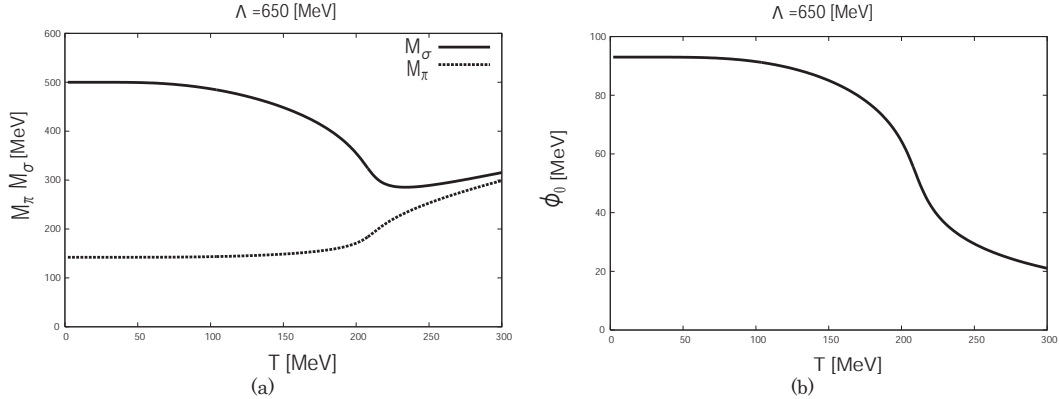}
\caption{The sigma and pion masses are shown as functions of temperature 
$T$ in (a). The order parameter of the chiral phase transition 
$\varphi_0$ is depicted in (b). 
The cutoff parameter $\Lambda$ is taken as 650 MeV.
}
\label{fig:4-1}
\end{center}
\end{figure}

In this section, we demonstrate the numerical results for the 
meson masses, $M_\pi$ and $M_\sigma$, and the mean filed value for 
$\varphi_0$ at finite temperature. 
The meson masses are given in Eqs.(\ref{2-29}) and (\ref{2-47}) 
with the propagator and polarization tensors replaced by 
Eqs.(\ref{3-3})$\sim$(\ref{3-5}) or (\ref{3-6})$\sim$(\ref{3-9}) 
in the finite temperature system. 
We have model parameters $m_0$, $\lambda$, $c$ and the three momentum cutoff 
$\Lambda$. 
We take these value to reproduce the pion mass 
(140 MeV), the sigma meson mass (500 MeV) and the condensate $\varphi_0$ 
(93 MeV) at zero temperature values in each value of $\Lambda$. 
Here, the three momentum cutoff $\Lambda$ is not determined in 
this framework. 
Therefore, we take several values of $\Lambda$.

It is well known that, in the Nambu-Jona Lasinio (NJL) model, 
the three momentum cutoff is taken as a rather small value 
around 630 MeV.\cite{HK}
Thus, we might have to adopt a rather small cutoff value. 
Under the above consideration, we first take the three momentum cutoff 
as 650 MeV. 
Thus, we neglect the polarization tensor $\Pi_{00}^{({\rm pol})}$ 
because this tensor represents the sigma meson pair excitation. 
We adopt the sigma meson mass at zero temperature as 500 MeV, 
so the pair has the energy 1000 MeV at least. 
This value is beyond the cutoff parameter $\Lambda$. 
This is the reason why we neglect the polarization tensor 
$\Pi_{00}^{({\rm pol})}$ in the following numerical calculation. 
Thus, we set $\Pi_{00}^{({\rm pol})}=0$ in the formula 
for the sigma meson mass, Eq.(\ref{2-47}). 

We show the pion and sigma meson masses and the mean field value 
$\varphi_0$ in Fig.\ref{fig:4-1} (a) and (b), respectively. 
The horizontal axis represents the temperature. 
In this case, the order parameter, ${\varphi_0}$, of the chiral phase 
transition monotonically decreases and is a single-valued function of 
the temperature $T$, 
so the order of phase transition may be likely the second order or 
crossover. 
Correspondingly, the pion mass and the sigma meson mass are also 
monotonically changed and they are single-valued function of $T$.

\begin{figure}[t]
\begin{center}
\includegraphics[height=5.3cm]{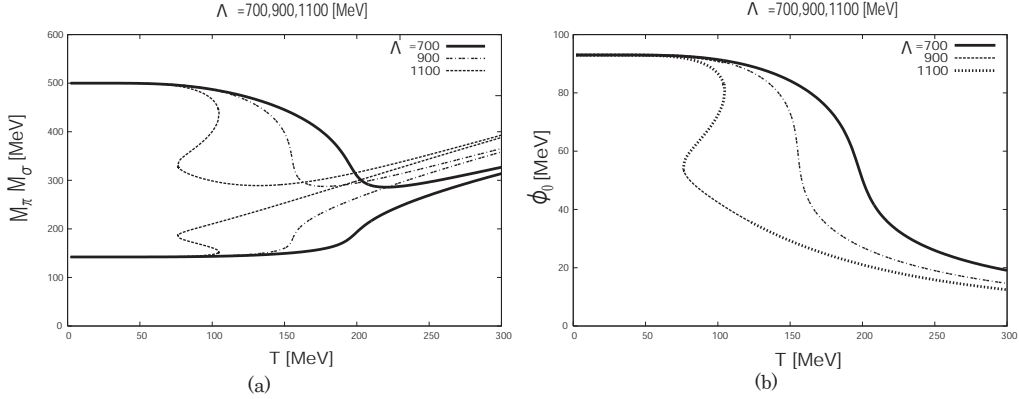}
\caption{The sigma and pion masses are shown as functions of temperature 
$T$ in (a). The order parameter of the chiral phase transition 
$\varphi_0$ is depicted in (b). 
The cutoff parameters $\Lambda$ are taken as 700 MeV (solid curves), 
900 MeV (dash-dotted curves) and 1100 MeV (dotted curves), respectively.
}
\label{fig:4-2}
\end{center}
\end{figure}
\begin{figure}[t]
\begin{center}
\includegraphics[height=5.3cm]{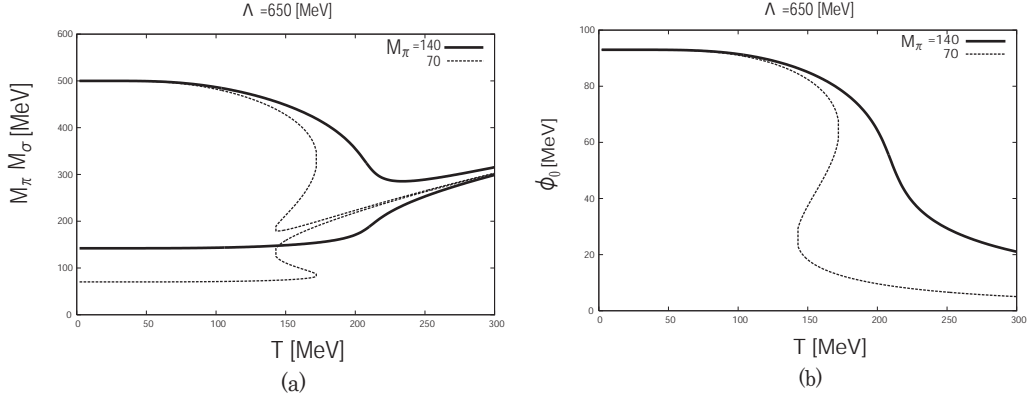}
\caption{The sigma and pion masses are shown as functions of temperature 
$T$ in (a). The order parameter of the chiral phase transition 
$\varphi_0$ is depicted in (b). 
The pion masses at zero temperature are taken as 140 MeV (solid curves) 
and 70 MeV (dotted curves), respectively. 
The cutoff parameter $\Lambda$ is taken as 650 MeV.
}
\label{fig:4-3}
\end{center}
\end{figure}

In Fig.\ref{fig:4-2}, we show the results of the pion mass, sigma meson mass 
and the mean field value of the order parameter $\varphi_0$ for the various 
values of the three momentum cutoff parameter $\Lambda$. 
Solid curves, dash-dotted curves and the dotted curves show 
the numerical results for the cutoff parameters $\Lambda=700$ MeV, 
900 MeV and 1100 MeV, respectively. 
If the cutoff parameter $\Lambda$ is taken as a rather larger value, 
the change of the order parameter is steep in the transition region. 
Beyond a certain value for $\Lambda$, the order parameter becomes 
a multi-valued function of $T$. 
Also, the meson masses become multi-valued 
functions of $T$. 
This behavior reveals the first order phase transition. 
Thus, it should be noted that the choice of the value of cutoff includes 
subtle problem such as the determination of 
the order of phase transition. 
It may be concluded 
that the rather small value for the cutoff $\Lambda$ should 
be adopted in this model, because the chiral phase transition 
at finite temperature may be crossover in the realistic parameterization 
for the pion mass. 

In Fig.\ref{fig:4-3}, the dotted curves show the results 
for the pion mass and sigma meson mass in (a) and 
the order parameter in (b) with the tree momentum cutoff $\Lambda=650$ MeV, 
where the pion mass at zero temperature 
is taken as 70 MeV unrealistically. 
The solid curves represent the results under the realistic value 
for zero temperature pion mass for the comparison.  
For lower pion mass, the order parameter becomes a multi-valued function 
of the temperature $T$, so the chiral phase transition becomes to 
the first order one. 
Thus, in the case of the chiral limit, the chiral phase transition 
is the first order one 
within our approach in this O(4) linear sigma model. 
This behavior is the same as that derived in Ref.\citen{CH}, where 
the optimized perturbation theory is used in the same model.

\section{Summary and concluding remarks}

In this paper, we investigated the chiral phase transition 
at finite temperature in the O(4) linear sigma model 
based on the time-dependent variational method devised by the 
linear response theory for the external fields. 
In this approach, the Goldstone theorem for the spontaneous chiral 
symmetry breaking is retained although the variational method 
is adopted. 
The reason why the above-mentioned situation is realized is that the 
meson-pair excitations are automatically included in this approach 
such as the particle-hole pair excitations for the collective 
motions of the nuclear many-body problem in the random phase 
approximation (RPA). 
It is well known that the pion as a Nambu-Goldstone boson has properties 
of the collective mode for the quark-antiquark pair 
in terms of the model based on the quarks.\cite{Weise}

In the O(4) linear sigma model, the nature of the collective mode is revealed 
with the pion-sigma meson pair excitations like the RPA in our variational 
approach devised by the linear response theory. 
The same situation is realized for the sigma meson which is the chiral partner 
of the pion in terms of the chiral symmetry. 
It was shown that the sigma meson contains the effects of 
the pion-pion and sigma meson-sigma meson pair excitations.

We extended this variational method to the finite temperature case 
by means of the imaginary time formalism for the finite temperature 
field theory. 
This is a natural extension in order to 
investigate the chiral phase transition at finite temperature. 

We have one free model parameter that is the three momentum 
cutoff parameter $\Lambda$ since we regarded the O(4) linear sigma model 
as a low energy effective model of QCD. 
As is similar to the NJL model, we adopted a rather small cutoff $\Lambda$ 
around 700 MeV. 
In this parameterization, the order parameter of the chiral phase transition 
is monotonically changed and is the single-valued function 
of the temperature $T$. However, the change of the order parameter 
in the transition region becomes rapid as the cutoff $\Lambda$ is larger. 
Beyond a certain value of $\Lambda$, the order parameter becomes 
a multi-valued function of $T$. 
Thus, it should be noted that the cutoff has to be determined carefully. 
The similar caution should be given for the pion mass. 
It was shown that, 
even if the cutoff parameter is taken as a small value, the order 
of the chiral phase transition changes according with the pion mass at zero 
temperature. 
Thus, the important problem may be to determine the value of cutoff 
through the physical process, for example, low energy pion-pion scattering, 
if the O(4) linear sigma model is used as a low energy 
effective model of QCD. 
Further, it is well known that a certain susceptibility presents 
an important information for the chiral phase transition. 
These investigations are future problems.

Recently, the chiral phase transition at finite temperature is 
investigated in the same model by using a slight different 
manner,\cite{Matsuo} 
in which 
the meson excitations are treated by introducing the Wigner functions 
and the Vlasov equations are formulated for these Wigner functions.\cite{MM} 
Thus, as another future problem, 
it may be important to understand the dynamical chiral phase 
transition in the several situations, for example, 
in the quench scenario,\cite{IAT} in the context of the 
high energy heavy ion collisions.

\section*{Acknowledgements} 
The authors would like to express their sincere thanks to Professors 
M. Iwasaki and K. Iida and the members of Hadron Physics 
Group of Kochi University for discussing the subjects in this 
paper and 
giving them valuable comments. 
One of the authors (Y.T.) also would like to express his thanks to Professor 
T. Matsui for the collaboration of the study in the variational 
approach to the O(N) linear sigma model. He also acknowledges to 
Professor Dominique Vautherin for the collaboration and giving him 
the suggestion for this work developed in this paper. 
He is partially supported by the Grants-in-Aid of the Scientific Research 
No.18540278 from the Ministry of Education, Culture, Sports, Science and 
Technology in Japan.


\end{document}